\documentclass[unsortedaddress,prd,preprint,showpacs]{revtex4-1}

\usepackage{amsmath}
\usepackage{amsfonts}
\usepackage{amsthm}
\usepackage{amssymb}
\usepackage{graphicx}
\usepackage{hyperref}

	\newcommand{\beq}{\begin{equation}}
	\newcommand{\eeq}{\end{equation}}

\begin{document}

	\title{Topological quantization of gravitational fields as generalized harmonic maps}
	\date{\today}

	\author{Francisco Nettel}
	\email{Francisco.Nettel@roma1.infn.it}
	\affiliation{Dipartimento di Fisica, Universit\`a di Roma La Sapienza,\\ P.le Aldo Moro 5, I-00185 Rome, Italy}

	\begin{abstract}
 	We apply the topological quantization method to some gravitational fields which can be represented as generalized harmonic maps. This representation extends the well-known concept of harmonic maps and allows us to describe some solutions to the Einstein field equations as generalized strings, where the base space for the generalized harmonic map is a two-dimensional manifold. The results obtained here point to an incompatibility of the classical description of the gravitational field and its symmetries with a discretization of the parameters that enter in its description. 
	\end{abstract}


\maketitle

\section{Introduction}  
\label{sec:intro}

Topological quantization (TQ) is a method inspired by Dirac's quantization of the charge in the presence of a magnetic monopole \cite{Dirac1936}. There, it is discussed the phase factor acquired by the wave function of a charged particle moving around a monopole configuration leading to a discrete relation between the charges of the monopole and the electron. Wu and Yang in \cite{WuYang} discussed this quantization condition and others aspects of gauge fields in terms of nonintegrable phase factors and the gauge transformations that relate them in the different overlapping regions of spacetime. The mathematics involved in their description is related to the theory of fiber bundles. The topological/geometric ideas upon which this quantization method relies have been developed and applied to monopole and instanton configurations \cite{schwarz, frankel,zhong} and similar analyses have been done in the context of cohomology \cite{alvarez}, topological models of electromagnetism \cite{bulgadaev} and current quantization \cite{ranada}. 

The motivation for this analysis comes mainly for the need to explore alternative options to face the problem of quantization of physical systems and in particular the more elusive problem of quantizing the gravitational field. Among the candidates to which more effort has been devoted we find loop quantum gravity \cite{rovelliLR} and superstring theory \cite{horowitz, blau}. In spite of the success that different approaches have reached, there are still problems which remain to be solved \cite{kiefer}. A major obstacle to the development of a quantum description of gravity is the lack of experimental evidence that can lead the way to a theoretical model. For this reason it is of great relevance the insight obtained from the phenomenology of quantum spacetime \cite{amelinoLR} which could give valuable information to help in the construction of different approaches. There are other approaches which have been growing recently, for instance, field theories with group momentum spaces are of interest in the spin foam approach to quantum gravity, in the analysis of relativistic particles coupled to 3d gravity and in studying point particles coupled to Chern-Simons description of 3d gravity \cite{meusburger, arzano1, arzano2}. It is significant that results from 3d quantum gravity which point to a curved momentum space and a noncommutativity of the spacetime coordinates can be understood as a theory with (DSR) deformed relativistic symmetries \cite{GAC1}, thus preserving an observer independent description of phenomena. 

In the method of topological quantization the analysis of the physical systems is based on a geometric structure which is a principal fiber bundle representing the system and its symmetries, considered as classical symmetries of spacetime. At this point in the program of TQ we expect to obtain quantum information in the form of discrete relations between the parameters that enter in the description of the physical system as it is done for gauge field theories in \cite{WuYang}, without imposing any extra conditions to those proper to the definition of the principal fiber bundle that represents the system. Therefore it is interesting for us to compare the results obtained here with the 
ideas of some approaches to quantum gravity, which in general lead to the notion of discretization of spacetime and deformation of symmetries. So far, TQ has been useful to obtain some topological spectra for different configurations, in particular the analysis of gravitational configurations \cite{QuePat1} which was the starting point of this program. Moreover, conservative mechanical systems \cite{yo1,nettel2} and bosonic strings \cite{Nettelstrings,gus2} have also been considered within this method. Recently, some gravitational configurations have been studied using as a standard fiber for the principal bundle the Lorentz group as the symmetry group (reducing from the group of diffeomorphisms). However, these preliminary results are still under discussion \cite{Quevedonew}. 

The outline of the paper is the following. In section \ref{sec:ghm}  we recall the generalized harmonic maps (GMH) defined in \cite{Nettelghm}, as an equivalent description for gravitational fields which possess two commuting Killing vector fields. The introduction of GHM originally pursued to describe these solutions to the gravitational field as generalized string models, something which in principle would allows us to canonically quantize the systems given that we know the solutions to the embedding that defines the generalized Polyakov action. The description of gravitational fields as GHM gives us an alternative to implement the method of topological quantization on a two-dimensional base space which in particular allows for a reduction of the spacetime symmetries to an Abelian group $SO(2)$ or $SO(1,1)$ depending of the signature of the induced metric on the base manifold. In section \ref{sec:tq} we present the method of topological quantization for the case of GHM and establish how to obtain the quantization conditions, if any. We include some examples which cover different options for gravitational fields including the Schwarzschild solution, the unpolarized Gowdy $T^3$ cosmological solution and the Einstein-Rosen gravitational waves. In all the cases examined here, we find that no quantization conditions are imposed. This result at first look seems surprising, but after some reflection we can start to ponder if there is a role for local classical symmetries in the quantum description of spacetime. It seems that our results point in the direction of modifying the fundamental symmetries of the gravitational field and considering the Lorentz group only as an effective local symmetry in the description of gravity in the classical regime \cite{Quevedonew}. In the last section \ref{sec:remarks} we present the conclusions.

\section{Generalized harmonic maps}
\label{sec:ghm}

In this section we present the generalization to the concept of harmonic maps \cite{Fuller,Misner} as is given in \cite{Nettelghm}. First, we recall the definition of harmonic maps. Let $(\mathcal{M},g)$ and $(\mathcal{N},G)$ be two (pseudo) Riemannian manifolds of dimension $m$ and $n$ and with local coordinates $x^a$, $a=1,\ldots, m$ and $y^A$, $A=1,\ldots,n$ respectively. A smooth map $f:\mathcal{M} \to \mathcal{N}$ is said to be harmonic if it satisfies the equations of motion obtained from the variation of the following action
	\beq  \label{harmap}
	S = \int d^m x\ L = \int d^m x\ \sqrt{|g|} g^{ab}(x) h_{ab}(x),
	\eeq
where $h_{ab} = \partial_a y^A \partial_b y^B G_{AB}(y)$ is the induced metric on the embedded manifold $f(\mathcal{M})$ and the mapping $f$ is represented in local coordinates as $f \to (y^1(x),\ldots,y^n(x))$. We shall call $\mathcal{M}$ the base space and $\mathcal{N}$ the target space. The equations of motion satisfied by the harmonic maps can be obtained from the variation of \eqref{harmap} with respect to the \textit{fields} $y^A(x)$ yielding
	\beq \label{eqharmap}
	\frac{1}{\sqrt{|g|}} \partial_a \big(|g|\ g^{ab} \partial_b y^A \big) + \Gamma^{A}_{\ BC}(y) g^{ab} \partial_a y^B \partial_b y^C = 0,
	\eeq
where $\Gamma^A_{\ BC}(y)$ are the Christoffel symbols associated to the metric of the target space $G$. We can define the energy-momentum tensor of the embedded hypersurface as $T_{ab} =  \frac{\delta L}{\delta g^{ab}}$, where $L$ is the Lagrangian density in \eqref{harmap}. 

The generalization introduced in \cite{Nettelghm} was originally intended to describe some particular gravitational configurations as some sort of generalized bosonic strings, that is, having as the base space of the mapping a two dimensional manifold which can be understood as a worldsheet embedded in the target space $\mathcal{N}$. In this work, we are interested in the particular feature of the GHM of having as base space $\mathcal{M}$ a two dimensional manifold. The generalized harmonic mapping is achieved allowing the target space metric to depend explicitly on the base space variables, $G=G(y,x)$. This modification can be interpreted as an interaction between the two spaces and as we will see, this modifies the conservation law for the energy-momentum tensor. The equations of motion for the GHM can be obtained from the variation with respect to $y^A$ of the action 
	\beq  \label{actionghm}
	S = \int d^m x\ L = \int d^m x\ \sqrt{|g|} g^{ab}(x) \partial_a y^A\ \partial_b y^B\ G_{AB}(y,x),
	\eeq
	where we must notice the explicit dependence of $G$ on the base space coordinates $G=G(y,x)$. Again we have the induced metric given by the generalized harmonic mapping $h = f^*(G)$ with components 
	\beq  \label{induced}
	h_{ab} = \partial_a y^A\ \partial_b y^B\ G_{AB}(y,x).
	\eeq
	
Then, the set of equations for this generalized mapping are
	\beq \label{eqghm}
	\frac{1}{\sqrt{|g|}} \partial_a \left(|g|\ g^{ab} \partial_b y^A \right) + \Gamma^{A}_{\ BC}(y,x) g^{ab} \partial_a y^B \partial_b y^C + G^{AB}(y,x) g^{ab}\ \partial_a y^C \partial_b G_{BC}(y,x) = 0,
	\eeq
where we have written explicitly the dependence on the $x^a$ coordinates for the relevant quantities. Obviously, the Lagrangian density and field equations reduce to the harmonic mapping when $\partial_a G_{AB} = 0$. 
Both harmonic mappings and their generalization are invariant under reparametrizations of the base and diffeormorphisms on the target space. In the particular case of a two-dimensional base space there is also invariance of the action under Weyl transformations. For a detailed analysis of these symmetries see the appendix in \cite{Nettelghm}. As for the conservation law, we have for the regular harmonic maps that $\nabla_b T^{ab} = 0$, while the relation for the GHM case is generalized as
	\beq \label{consghm}
	\nabla_b T_a^{\ b} + \frac{1}{2}\frac{\partial L}{\partial x^a} = 0,
	\eeq
where both conservation laws are understood to be on-shell with their corresponding field equations. We will find that relation \eqref{consghm}, for the gravitational fields we study in this work, carry information about the gravitational function which is solved by quadratures once the main Einstein field equations are solved, which otherwise is lost when the system is described by the reduced Lagrangian. For a more detailed analysis of the properties of GHM see \cite{Nettelghm}.

In the following subsections we will use GHM to describe gravitational fields which possess two commuting Killing vector fields. In particular, we will deal with stationary axisymmetric solutions along with the limit case of static spacetimes, the unpolarized Gowdy $T^3$ spacetime and Einstein-Rosen gravitational waves.

\subsection{Stationary axisymmetric spacetimes}
\label{sub:static}

The line element for stationary axisymmetric spacetimes in Weyl-Lewis-Papapetrou coordinates is given by
	\beq  \label{dsaxi}
	ds^2 = -f(dt - \omega d\phi)^2 + \frac{1}{f} \bigg[e^{2 \kappa} (dz^2 + d\rho^2) + \rho^2 d\phi^2 \bigg],
	\eeq
where $f$, $\kappa$ and $\omega$ are functions of $\rho$ and $z$. The field equations are
	\beq \label{eqmain1}
	\partial^2_\rho f + \frac{1}{\rho} \partial_\rho f + \partial^2_z f - \frac{1}{f}\bigg[(\partial_\rho f)^2 + (\partial_z f)^2 - (\partial_\rho \Omega)^2 - (\partial_z \Omega)^2 \bigg] = 0, 
	\eeq
	\beq \label{eqmain2}
	\partial^2_\rho \Omega + \frac{1}{\rho} \partial_\rho \Omega + \partial^2_z \Omega - \frac{2}{f} \bigg[\partial_\rho f \partial_\rho \Omega + \partial_z f \partial_z \Omega \bigg] =0,
	\eeq
	\beq  \label{cuad1}
	\partial_\rho \kappa - \frac{\rho}{4 f^2}\bigg[(\partial_\rho f)^2 - (\partial_z f)^2 + (\partial_\rho \Omega)^2 - (\partial_z \Omega)^2 \bigg] =0,
	\eeq 
	\beq  \label{cuad2}
	\partial \kappa_z - \frac{\rho}{2 f^2} \bigg[\partial_\rho f \partial_z f + \partial_\rho \Omega \partial_z \Omega \bigg] = 0,
	\eeq
where we have introduced the function $\Omega(\rho, z)$ as
	\beq \label{Omega}
	\partial_\rho \omega = -\frac{\rho}{f^2} \partial_z \Omega \qquad \text{and} \qquad \partial_z \omega = \frac{\rho}{f^2} \partial_\rho \Omega.
	\eeq

The two first equations \eqref{eqmain1} and \eqref{eqmain2} are known as the main field equations as they determine the complete solution. In fact, solving the main field equations for $f$ and $\Omega$, then the other pair of equations \eqref{cuad1} and \eqref{cuad2} give the solution $\kappa$ by cuadratures. It is possible to obtain the main field equations from the reduced Lagrangian density \cite{Ernst1}
	\beq  \label{predlagaxi}
	L = \frac{\rho}{2 f^2} \bigg[ (\partial_\rho f)^2 + (\partial_z f)^2 \bigg] + \frac{f^2}{\rho} \bigg[ (\partial_\rho \omega)^2 + (\partial_z \omega)^2) \bigg],
	\eeq 
which in terms of the alternative field $\Omega$ is expressed as
	\beq	  \label{redlagaxi}
	L = \frac{\rho}{2 f^2} \bigg[ (\partial_\rho f)^2 + (\partial_z f)^2 + (\partial_\rho \Omega)^2 + (\partial_z \Omega)^2 \bigg],
	\eeq
for the action integral
	\beq \label{redaction}
	S = \int_\mathcal{M} L\  dt d\rho d\phi dz ,
	\eeq
where $\mathcal{M}$ is the spacetime manifold.

The Lagrangian density \eqref{predlagaxi} can be obtained from the curvature scalar neglecting the total divergence terms and noticing that $\omega$ and $\kappa$ are ignorable fields. In this way, we can arrive at its final form using the Routh procedure \cite{Goldstein}. This reduced Lagrangian density can also be interpreted as the Lagrangian for the non-linear sigma model $SL(2,R)/SO(2)$. We will not pursue this construction here, for a detailed analysis see \cite{Cortes1}. In terms of harmonic maps is constructed with the 4-dimensional spacetime with the metric associated to the line element \eqref{dsaxi} and a two-dimensional target space with coordinates $y^A = f, \Omega$ and a diagonal metric $G = \tfrac{1}{2f^2} \text{diag}(1,1)$. Then, from the Lagrangian density in \eqref{harmap} and the field equations \eqref{eqharmap} we can recover the reduced Lagrangian \eqref{redlagaxi} and the main field equations \eqref{eqmain1} and \eqref{eqmain2}, respectively. 

As we mentioned before, we would like to construct a representation for the gravitational fields whose base space is a two-dimensional manifold (basically because the symmetries are reduced to that described by an Abelian group and in principle the analysis could be easier). As can be seen from \eqref{redlagaxi}, the reduced Lagrangian depends explicitly on the base space coordinates. In this case and for those in the following paragraphs, this property prevents us from using harmonic maps to describe the gravitational configurations as mappings from a two-dimensional base space to a 2-dimensional target space which reproduce the relevant Lagrangian \eqref{redlagaxi} and the field equations \eqref{eqmain1} and \eqref{eqmain2}. Introducing the GHM permits us to overcome this difficulty and still have a consistent geometric picture, with the bonus feature of recovering the equations \eqref{cuad1} and \eqref{cuad2} by means of the generalized conservation relations \eqref{consghm}, otherwise lost when obtaining the reduced Lagrangian.

The representation through GHM is accomplished once the metrics for the base and target spaces are provided. By simple inspection of the Lagrangian density \eqref{redlagaxi} we realize that the appropriate metrics are
	\beq \label{axighmetrics}
	g_{ab} = \delta_{ab} \qquad \text{and} \qquad G_{AB} = \frac{\rho}{2 f^2} \delta_{AB},
	\eeq
where $x^a = (\rho,z)$ are the coordinates in $\mathcal{M}$ and $y^A = (f, \Omega)$ in $\mathcal{N}$. We notice in the metric for $\mathcal{N}$ the explicit dependence of the metric components on the $\mathcal{M}$ coordinates, $G = G(f,\Omega, \rho)$. It is remarkable that the components of the energy-momentum tensor can be expressed in terms of the metric function $\kappa(\rho,z)$ using the equations (\ref{cuad1}, \ref{cuad2})
	\beq  \label{axiemrr}
	T_{\rho\rho} = -T_{zz} = \partial_\rho \kappa \qquad,
	\eeq
and
	\beq  \label{axiemrz}
	T_{\rho z} = \partial_z \kappa.
	\eeq

The generalized conservation law \eqref{consghm} for the $\rho$ component is
	\beq  \label{axiconsr}
	\frac{dT_{\rho\rho}}{d\rho} + \frac{dT_{\rho z}}{dz} + \frac{1}{2} \frac{\partial{L}}{\partial \rho} = 0,
	\eeq
where of course it is understood that the relation is fulfilled on-shell and the partial differentiation acts only on the explicit dependence on $\rho$. For the $z$ component we have 
	\beq  \label{axiconsz}
	\frac{dT_{z\rho}}{d\rho} = - \frac{dT_{zz}}{dz},
	\eeq
which is recognized as the integrability condition for the partial differential equations (\ref{cuad1},\ref{cuad2}), $\partial_{\rho z} \kappa = \partial_{z \rho} \kappa$. It is interesting that the information for the metric function $\kappa$ neglected in the reduced Lagrangian density is present in the form of a generalized conservation law for the energy-momentum tensor. Therefore, we claim that this geometric representation through GHM is completely equivalent to the solution starting from the Einstein-Hilbert action.

\subsubsection*{Dimensional extension for the static solution} \label{subsub:static}

The particular case of a static spacetime is obtained if we set $\Omega = 0$. In the context of GHM this becomes a degenerate case given that the target manifold $\mathcal{N}$ turns into a one-dimensional manifold and no non-degenerate metric can be associated to it. In order to include the static case within the GHM description we can define a dimensional extension for $\mathcal{N}$ as follows. Let us separate the degrees of freedom in the target space by means of the coordinatization $(y^A, y^{\tilde{A}})$, where $y^A$ will contain all the information pertaining to the gravitational field, whereas $y^{\tilde{A}}$ will describe the dimensional extension sector. In order to not disrupt the field equations which already describe correctly the gravitational field we will ask for the target space metric $G$ to comply with the following: $G_{A\tilde{B}} = 0$, $\partial_{\tilde{A}} G_{AB} =0$ and $\partial_A G_{\tilde{A}\tilde{B}} = 0$. It is possible to show that the Lagrangian density \eqref{actionghm} gets an additional the term $\tilde{L} = G_{\tilde{A}\tilde{B}} (\partial_\rho y^{\tilde{A}} \partial_\rho y^{\tilde{B}} + \partial_z y^{\tilde{A}} \partial_z y^{\tilde{B}})$, yielding a set of equations for the extension sector
	\beq \label{eqdimext}
	\partial^2_\rho y^{\tilde{A}} + \partial^2_z y^{\tilde{A}} + \Gamma^{\tilde{A}}_{\ \tilde{B}\tilde{C}} \big( \partial_\rho y^{\tilde{B}} \partial_\rho y^{\tilde{C}} + \partial_z y^{\tilde{B}} \partial_z y^{\tilde{C}} \big) + G^{\tilde{A} \tilde{C}} \big( \partial_\rho y^{\tilde{B}} \partial_\rho G_{\tilde{B}\tilde{C}} + \partial_z y^{\tilde{B}} \partial_z G_{\tilde{B}\tilde{C}} \big) = 0.
	\eeq
This extension permits us to describe the static solutions as GHM choosing the minimum extension of a two-dimensional target space with coordinates $(y^A = f, y^{\tilde{A}} = X)$ and metric $G = \text{diag}(\tfrac{\rho}{2 f^2}, 1)$. The only main field equation is given by \eqref{eqmain1} setting $\Omega = 0$ and the equations for $\kappa$ are (\ref{cuad1},\ref{cuad2}) also with $\Omega = 0$, while the equation for the extension is described in terms of a harmonic function for $y^{\tilde{A}}$
	\beq  \label{ytilde}
	(\partial^2_\rho + \partial^2_z) y^{\tilde{A}} = 0.
	\eeq

So far we have been working in the Weyl coordinates but in order to describe the Schwarzschild solution it will be convenient to  switch to a coordinate system adapted to this more symmetric spacetime, namely $(t,r,\theta,\varphi)$. Constructing the reduced Lagrangian by eliminating the second order terms in the derivatives for the general form of the line element
	\beq \label{schwforma}
	ds^2 = -f(r) dt^2 + f^{-1}(r) dr^2 + r^2 ( d\theta^2 + \sin^2\theta d\varphi^2),
	\eeq
we find a linear first order expression in the derivatives of the field $f(r)$ which obviously cannot be expressed as a GHM, given that we need quadratic terms in the derivatives and  besides the target space becomes one-dimensional. Using the dimensional extension we can write down a generalized Lagrangian density which does the trick without modifying the gravitational sector of the mapping starting from a two-dimensional base space with coordinates $(r,\varphi)$ where $\varphi$ is chosen as the second coordinate for convenience. The reduced Lagrangian is
	\beq  \label{gmhlagschw}
	L = r^2 \bigg(\frac{df}{dr} \bigg)^2 + r \left(\partial_r X\right)^2 + r^{-1} \left( \partial_\varphi X\right)^2.
	\eeq 
The GHM is determined by providing the metrics. For the two-dimensional base space $\mathcal{M}$ with coordinates $(r,\varphi)$ we have
	\beq  \label{basemetricschw}
	g = \text{diag}(1,r^2),
	\eeq
while for the target space $\mathcal{N}$ with coordinates $(f,X)$ we require that
	\beq  \label{targetmetricschw}
	G = \text{diag}(r, 1).
	\eeq

This is the minimum choice we can make for both spaces in order to reproduce the Lagrangian \eqref{gmhlagschw}. The field equations for the GHM are
	\beq  \label{eqsghmschw}
	\frac{d^2 f}{dr^2} + \frac{2}{r} \frac{df}{dr} = 0,
	\eeq
and
	\beq  \label{eqsghmschwext}
	\partial^2_r X + \frac{1}{r}\partial_r X + \frac{1}{r^2} \partial^2_\varphi X = 0,
	\eeq
from which we obtain the usual equation for $f$ and we observe that the field from the dimensional extension is any harmonic function $X$.

In the following subsections we describe the cases of unpolarized Gowdy $T^3$ spacetime and Einstein-Rosen gravitational waves. As many aspects of the description are completely analogous we will be more brief.

\subsection{Gowdy unpolarized $T^3$}
\label{sub:gowdy}

The line element that describes this class of spacetimes is 
	\beq  \label{dsGowdy}
	ds^2 = - e^{\frac{-\lambda + 3t}{2}} dt^2 + e^{-t + P}(d\sigma + Qd\delta)^2 + e^{-t - P} d\delta^2 + e^{\frac{-\lambda + t}{2}} d\theta^2,
	\eeq
where $(t,\sigma,\delta,\theta)$ are coordinates for spacetime and the metric functions $P$, $Q$ and $\lambda$ only depend on $(t,\theta)$. This class of vacuum solutions are interpreted as cosmological models which are closed and present inhomogeneities. They present a singularity at $t \to \infty$ so they can be used to explore some properties of an initial cosmological singularity. 
The reduced Lagrangian density is given by
	\beq \label{redlagGowdy}
	L = \frac{1}{2} \bigg[(\partial_t P)^2 - e^{-2t} (\partial_\theta P)^2 \bigg] + \frac{e^{2P}}{2} \bigg[ (\partial_t Q)^2 - e^{-2t} (\partial_\theta Q)^2 \bigg],
	\eeq
where $\lambda$ has been removed by similar arguments as used in the previous case. The main field equations are
	\beq  \label{eqmainGowdy1}
	\partial^2_t P - e^{-2t} \partial^2_\theta P - e^{2P} \big[ (\partial_t Q)^2 - e^{-2t} (\partial_\theta Q)^2 \big] = 0,
	\eeq
	\beq  \label{eqmainGowdy2}
	\partial^2_t Q - e^{-2t}\partial^2_\theta Q + 2 \left( \partial_t P \partial_t Q - e^{-2t} \partial_\theta P \partial_\theta Q \right) =0,
	\eeq
while in order to determine the $\lambda$ function we have the quadrature equations,
	\beq  \label{cuad1Gowdy}
	\partial_t \lambda = (\partial_t P)^2 + e^{-2t} (\partial_\theta P)^2 + e^{2P} \left[ (\partial_t Q)^2 + e^{-2t} (\partial_\theta Q)^2 \right],
	\eeq
	\beq   \label{cuad2Gowdy}
	\partial_\theta \lambda = 2 \left( \partial_t P \partial_\theta P + e^{2P} \partial_t Q \partial_\theta Q \right].
	\eeq

The representation as GHM is obtained by equipping the base space with coordinates $(r,\theta)$ and with the metric
	\beq  \label{basemetricGowdy}
	g = \text{diag}(1,-e^{-2t}),
	\eeq
and the target space with coordinates $(P,Q)$ and metric
	\beq  \label{targetmetricGowdy}
	G = \frac{e^{-t}}{2} \text{diag}(1, e^{2P}).
	\eeq
The field equations (\ref{cuad1Gowdy}, \ref{cuad2Gowdy}) bear a similar relation with the energy-momentum tensor as we have seen in the case of stationary axisymmetric spacetimes. The class of Gowdy spacetimes which we will here consider are the so-called asymptotically velocity term dominated cosmologies, whose behavior near the intital singularity $t \to \infty$ is described by the metric functions \cite{gowdyavtd}
	\beq  \label{avtdP}
	P(t,\theta) = \ln \big[ A(e^{-Ct} + B^2 e^{Ct} \big],
	\eeq
	\beq  \label{avtdQ}
	Q(t,\theta) = \frac{B}{A (e^{-2Ct} + B ) } + D,
	\eeq
with $A$, $B$, $C$ and $D$ functions of $\theta$. Later we will work with a convenient choice of these functions.

\subsection{Generalized Einstein-Rosen gravitational waves}
\label{sub:ERwaves}

The line element which describes this kind of solutions in the coordinate system $(t,\rho,\phi, z)$ is given by
	\beq  \label{dsER}
	ds^2 =  e^{-2 \psi}\big[e^{2\kappa}(- dt^2 + d\rho^2) + \rho^2 d\phi^2 \big] + e^{2\psi} \left(dz + \omega d\phi \right)^2,
	\eeq
where the metric functions $\psi, \kappa, \omega$ depend only on the coordinates $(t,\rho)$. These solutions, as in the previous cases, possess two commuting Killing vector fields and can be interpreted as cylindrically symmetric gravitational waves which become singular at the axis of symmetry. In case that the Killing vector fields are spacelike hypersurface-orthogonal it is possible to set $\omega = 0$, the metric becomes diagonal and the gravitational waves are linearly polarized \cite{exactsolutions}. 

The field equations for the metric functions are
	\beq  \label{ERmain1}
	-\partial^2_t \psi + \partial^2_\rho \psi + \frac{1}{\rho}\partial_\rho \psi - \frac{e^{-4\psi}}{2} \bigg[ (\partial_t \Omega)^2 - (\partial_\rho \Omega)^2 \bigg] = 0,
	\eeq
	\beq  \label{ERmain2}
	-\partial^2_t \Omega + \partial^2_\rho \Omega + \frac{1}{\rho} \partial_\rho \Omega + 4 \left( \partial_t \psi \partial_t \Omega - \partial_\rho \psi \partial_\rho \Omega \right) = 0,
	\eeq
	\beq  \label{ERcuad1}
	\partial_t \kappa = 2\rho \partial_t \psi \partial_\rho \psi + \frac{\rho e^{-4\psi}}{2} \partial_t \Omega \partial_\rho \Omega = 0,
	\eeq
	\beq  \label{ERcuad2}
	\partial_\rho \kappa = \rho \big[ (\partial_t \psi)^2 + (\partial_\rho \psi)^2 \big] + \frac{\rho e^{-4\psi}}{4} \bigg[ (\partial_t \Omega)^2 + (\partial_\rho \Omega)^2 \bigg] = 0,
	\eeq
where we have introduced $\omega_\rho = \rho e^{-4\psi} \partial_t \Omega$ and $\omega_t = \rho e^{-4\psi} \partial_\rho \Omega$. The reduced Lagrangian from which we can obtain the main field equations (\ref{ERmain1},\ref{ERmain2}) is 
	\beq  \label{ERredlag}
	L = 2\rho \big[ (\partial_t \psi)^2 - (\partial_\rho \psi)^2 \big] + \frac{\rho e^{-4\psi}}{2} \bigg[ (\partial_t \Omega)^2 - (\partial_\rho \Omega)^2 \bigg].
	\eeq 
This can be represented as a GHM providing the base space $\mathcal{M}$ with coordinates $(t, \rho)$ and the target space $\mathcal{N}$ with coordinates $(\psi, \Omega)$ with the metrics $g$ and $G$, respectively, as follows
	\beq   \label{ERmetrics}
	g = \text{diag}(1,-1) \qquad \text{and} \qquad G = \frac{\rho}{2}\text{diag}(4, e^{-4\psi}).
	\eeq 	
Again, the information for $\kappa$ is recovered in terms of the generalized conservation law for the energy-momentum tensor of the two-dimensional manifold $T_{ab} = \delta L/\delta g^{ab}$. 

For more details on the representation of these gravitational fields as GHM and their interpretation as generalized string models we refer to \cite{Nettelghm}.

\section{The method of topological quantization}
\label{sec:tq}

The method of topological quantization is basically a proposal for extending Dirac's idea about charge quantization \cite{Dirac1936} to any physical system, in particular gravitational fields. This method is intended more on a geometrical and topological setting as was interpreted by Wu and Yang in \cite{WuYang}, than in terms of the phase acquired of a state function when moving a charged particle around a charged monopole. As in any physical theory, topological quantization should provide, at some point of its development, at least three elements in order to be complete: the observables, the states associated to particular values of the observables and their dynamical evolution. In this work we are concerned only with the first of these elements and we focus on understanding the role of the symmetries of spacetime within this approach. For  defining states and their dynamics there are preliminary ideas which are being studied altogether with the way in which one should understand the symmetries of spacetime in the context of this method \cite{Quevedonew}. 

The method of topological quantization requires, as a first step, to establish the geometric representation of physical systems in the form of a principal fiber bundle (pfb). In general terms we can describe this method as follows: the physical information of the system must be contained in the base manifold of the pfb, and in a connection defined on it. The structure group will encode the symmetries of the system, hence the symmetry group is isomorphic to the standard fiber. Given an atlas for the base manifold the compatibility conditions on the transition functions on the fibers over the non-empty intersections of the corresponding charts will determine the discretization conditions on the physical parameters which enter in the expressions for the connection. Of course, this is just the standard geometrical set-up used to describe gauge theories when working with the connection and the field intensity (curvature) on the base manifold by means of the pull-back induced by the local sections in the pfb. The aim of topological quantization is to extend these ideas to gravitational systems. 

In the following paragraphs we will present the method of topological quantization for obtaining observables' spectra according to \cite{QuePat1} and \cite{Nettelstrings}, and we will apply this method to the general case of GHM and in particular to the examples described in section \ref{sec:ghm}.

\subsection{Topological quantization of GHM}
\label{sub:tqghm}

As a first step we must individuate the base manifold for the pfb from the different options that appear in the GHM representation. It is not difficult to convince oneself that the appropriate option is the embedded manifold $f(\mathcal{M})$ in $\mathcal{N}$ together with the connection associated with the induced metric on it. We choose this manifold as the base for the construction of the principal fiber bundle because the embedding equations coincide with the field equations that determine the gravitational configuration. Moreover, the energy-momentum tensor for the embedded manifold contains the information regarding the cyclic field which disappears when reducing the Lagrangian density from the Einstein-Hilbert Lagrangian. As for the symmetries of the system we take as the structure group the reparametrization of the base space $x'^a = x'^a(x^b)$ for the generalized harmonic map, which inherits the role of the invariance under diffeomorphism of the four-dimensional usual representation from the Einstein-Hilbert action \cite{QuePat1}. Introducing an orthonormal frame on each point of $f(\mathcal{M})$ we can reduce this symmetry to the group $SO(2)$ or $SO(1,1)$ depending on the signature of the induced metric. Therefore we have as the base space the manifold $(f(\mathcal{M}),h)$, where $h$ is the induced metric, from which we can find the compatible connection one-form $\omega_h$ on $f(\mathcal{M})$ with values in the Lie algebra $so(2)$ (resp. $so(1,1)$) and a structure group $SO(2)$ (resp. $SO(1,1)$) isomorphic to the standard fiber. 

Therefore, we can state the following result: Given a physical system described as a generalized harmonic map $(2 \to D)$ we can represent it as a unique principal fiber bundle $\mathcal{P}$, with $(f(\mathcal{M}), h)$ as the base manifold, $SO(2)$ (resp. $SO(1,1)$) as the structure group isomorphic to the standard fiber and a connection one-form with values in the Lie algebra $so(2)$ (resp. $so(1,1)$). It is possible to prove the existence and uniqueness of $\mathcal{P}$ using the reconstruction theorem for principal fiber bundles showing that all the elements required for its construction are present, namely, a base space, a structure group and a family of transition functions identified with elements of the structure group which fulfil the cocycle condition on the non-empty intersection of the open sets of the charts covering the base manifold.

The existence of a connection on $\mathcal{P}$ is established by a well-known result \cite{koba} which in the case of a $SO(k,l)$ structure group with $k$ and $l$ any integers can be cast in the following way: Let us consider a collection of open sets $\{U_i\}$ that cover $\mathcal{M}$, a set of one-forms with values in the Lie algebra $so(k,l)$ satisfying the compatibility conditions 
	\beq  \label{compatible}
	\omega_i = \Lambda_{ij} \omega_j \Lambda^{-1}_{ij} + \Lambda_{ij} d \Lambda^{-1}_{ij},
	\eeq
where $\Lambda_{ij}: U_i \cap U_j  \to SO(k,l)$ and a set of local sections $\sigma_i: U_i \to \pi^{-1}(U_i)$ such that in any $U_i \cap U_j \neq \emptyset$ we have $\sigma_i = \sigma_j \Lambda_{ij}$ (the indexes $i, j$ refer to the corresponding open set). Then, there is a unique connection $\tilde{\omega}$ on $\mathcal{P}$ such that $\omega_i = \sigma^*_i \tilde{\omega}$ where the pullback $\sigma^*_i$ is induced by the local section $\sigma_i$. For a more detailed exposition of a scheme of the proof for this proposition see \cite{QuePat1, nettel2}. In the cases we are dealing with here, the structure groups will be identified with $SO(2)$ and $SO(1,1)$, reducing the condition \eqref{compatible} to
	\beq  \label{compatible2}
	\omega_i = \omega_j +  \Lambda_{ij} d \Lambda^{-1}_{ij},
	\eeq
as the two options are abelian. This considerably simplifies the analysis of the compatibility conditions as we will see later, and it is an advantage given by the representation of the gravitational fields as GHM.

Now, let us be specific about how to reduce the invariance under reparametrizations to the orthogonal group $SO(2)$ or $SO(1,1)$, which is the manifestation of the diffeomorphism invariance of the four-dimensional theory, according to our representation through GHM. We introduce at each point of $\mathcal{M}$ an orientable orthonormal frame $\{e_a\}$, $a=1,\ldots, \text{dim}\ \mathcal{M}$, with respect to the Riemannian induced metric $h(e_a,e_b) = \delta_{ab}$ (for the pseudo-Riemannian case $h(e_a,e_b) = \eta_{ab}$). Given another frame $\{e_{a'}\}$ the two frames are related by $e_{a'} = (\Lambda^{-1})^b_{\ a'} e_b$ with $\Lambda \in SO(k,l)$. Associated to this orthonormal frame there are the one-forms that constitute the dual basis $\{\Theta^a\}$,  such that $\langle \Theta^a, e_b \rangle = \delta^a_{\ b}$. The dual basis allows us to express locally the metric as $h = h_{\mu\nu} dx^\mu \otimes dx^\nu = \delta_{ab}\, \Theta^a \otimes \Theta^b$ (or $h = h_{\mu\nu} dx^\mu \otimes dx^\nu = \eta_{ab}\, \Theta^a \otimes \Theta^b$). The Cartan structure equations permit us to find the local expressions for the connection one-form $\omega$ and curvature two-form $R$ on $\mathcal{M}$, 
	\beq \label{cartan1}
	d\Theta + \omega \wedge \Theta = 0,
	\eeq
	\beq  \label{cartan2}
	R = d\omega + \omega \wedge \omega,
	\eeq
where $d$ is the exterior derivative and we have omitted the frame indexes in $\Theta$, $\omega$ and $R$. Both, the connection one-form and the curvature two-form take values in the Lie algebra $so(k,l)$, thus under a change of orthonormal frame $\Theta' = \Lambda \Theta$ they transform as 
	\beq  \label{transnormw}
	\omega' = \Lambda \omega \Lambda^{-1} + \Lambda d \Lambda^{-1},
	\eeq
	\beq   \label{transnormR}
	R' = \Lambda R \Lambda^{-1}.
	\eeq
In this way we construct a one-form $\omega_i$ for each open subset $U_i$ of the covering for $\mathcal{M}$ which takes values in the $so(k,l)$. Moreover, wherever $U_i \cap U_j \neq \emptyset$ the transformation \eqref{transnormw} gives us the compatibility condition \eqref{compatible}. It is in the construction of the principal fiber bundle where the condition \eqref{compatible} must be fulfilled for a connection to be well-defined upon a covering $\{U_i\}$ for $\mathcal{M}$. This in turn implies having a \textit{gauge} transformation $\Lambda$ that is regular and single-valued in the intersections. The physical parameters of the system naturally enter in the expression for the connection $\omega$, thence the conditions for $\Lambda$ to be well-defined will be expressed as relations among these parameters. These relations define the topological quantization spectra.

In the following we will apply this method to the gravitational fields described by means of the GHM and see what are the corresponding quantization conditions obtained.


\subsubsection{The Schwarzschild solution}
\label{ssub:schw}

The first case is that of static solutions, among which the Schwarzschild solution is the physically relevant case in the limit of static spacetimes from axially symmetric stationary spacetimes. To find the quantization condition first we have to calculate the induced metric on $f(\mathcal{M})$ given the pair of metrics for the base \eqref{basemetricschw} and target spaces \eqref{targetmetricschw} of the GHM. From \eqref{induced} we find the induced metric
	\beq  \label{indschw}
	h = r \bigg[ \left(\frac{d f}{dr} \right)^2 + (\partial_r X)^2  \bigg]\ dr \otimes dr + 2\ \partial_r X\, \partial_\theta X\ dr \otimes d\theta + (\partial_\theta X)^2 d\theta \otimes d\theta,
	\eeq
where $X(r,\theta)$ is the function introduced in the definition of the dimensional extension. The only condition we impose on $X(r,\theta)$ is that it must be a harmonic function, i.e., it must satisfy \eqref{eqdimext}, otherwise is arbitrary. We can reduce the invariance under diffeormorphism on $f(\mathcal{M})$ to the group $SO(2)$ by introducing an orthonormal frame $\{e_a\}$, $a=1,2$ related to the coordinate basis by
	\beq  \label{frameschw}
	e_1 = \frac{1}{\partial_\theta X}\, \partial_\theta  \qquad \text{and} \qquad e_2 = \frac{1}{\sqrt{r}}\, \left(\frac{df}{dr}\right)^{-1} \bigg( \partial_r - \frac{\partial_r X}{\partial_\theta X}\, \partial_\theta \bigg),
	\eeq
while the dual basis $\{\Theta^a\}$ is
	\beq  \label{dframeschw}
	\Theta^1 = \partial_r X\, dr + \partial_\theta X\, d\theta  \qquad \text{and} \qquad \Theta^2 = \sqrt{r}\, \frac{df}{dr}\, dr.
	\eeq
By a simple calculation using Cartan first structure equation \eqref{cartan1} we obtain that the connection one-form vanishes for any solution $f(r,\theta)$ and extension field $X(r,\theta)$. Therefore, no quantization conditions are imposed for static solutions of this form. In particular we find that \textit{there is not a discrete spectrum} for the mass parameter of the Schwarzschild gravitational field, which is the only physical quantity that enters in $f$. We will discuss this lack of discretization by trivial quantization conditions later. It must be said that preliminary results following the method for the four-dimensional representation from Einstein-Hilbert action for this case also show that no quantization conditions are imposed \cite{Quevedonew}. Before moving to the next case we would like to comment on something peculiar about this representation of the Schwarzschild solution. Given that the connection is identically zero we have that the curvature scalar also vanishes, then we see that the embedded manifold is flat and there is no longer any a trace of the curvature singularity within this reduced representation. It is clear that, although this representation by GHM is completely equivalent in the sense that it contains all the information (field equations, conservation laws), one must be careful in interpreting the embedded manifold in relation with the four-dimensional spacetime manifold. One should always rely on this latter manifold to understand the classical spacetime.


\subsubsection{The unpolarized Gowdy solution}
\label{ssub:tqgowdy}

Here we study the case of the unpolarized Gowdy $T^3$ solution. The induced metric is obtained given the metrics of the GHM \eqref{basemetricGowdy} and \eqref{targetmetricGowdy}
	\begin{multline}  \label{indmetGowdy}
	h = \frac{e^{-t}}{2}  \bigg( e^{2P} (\partial_t Q)^2 + (\partial_t P)^2 \bigg) dt \otimes dt + e^{-t}\left( e^{2P} \partial_t Q \partial_\theta Q + \partial_t P \partial_\theta P \right) dt \otimes d\theta \\ 
	+ \frac{e^{-t}}{2} \bigg( e^{2P} (\partial_\theta Q)^2 + (\partial_\theta P)^2 \bigg) d\theta \otimes d\theta.
	\end{multline}
The signature of this metric is Euclidean, thus introducing an orthonormal frame $\{e_a\}$ we reduce to the local symmetry to that of $SO(2)$ for the GHM  
	\begin{align}  \label{framegowdy}
	e_1 &= \frac{\sqrt{2}}{\sqrt{ e^{-t} (e^{2P} (\partial_t Q)^2 + (\partial_t P)^2)}} \partial_t,  \\
	e_2 &= \frac{\sqrt{2} e^{t- P} \sqrt{2 e^{-t} (e^{2P} (\partial_t Q)^2 + (\partial_t P)^2)}}{ \partial_\theta Q \partial_t P - \partial_t Q \partial_\theta P}\bigg( \frac{e^{-t} (e^{2P} \partial_t Q \partial_\theta Q + \partial_t P \partial_\theta Q)}{\sqrt{ e^{-t} (e^{2P} (\partial_t Q)^2 + (\partial_t P)^2)}} \partial_t - \partial_\theta \bigg),
	\end{align}
while the dual basis is
	\begin{align} \label{dframegowdy}
	\Theta^1 &= \frac{1}{\sqrt{2}}\sqrt{e^{-t} (e^{2P} (\partial_t Q)^2 + (\partial_t P)^2)}\left( dt + \frac{e^{2P} \partial_t Q \partial_\theta Q + \partial_t P \partial_\theta P}{e^{2P} (\partial_t Q)^2 + (\partial_t P)^2} d\theta \right),  \\
	\Theta^2 &= \frac{e^{-t +P}}{\sqrt{2}}\frac{\partial_t P \partial_\theta Q - \partial_t Q \partial_\theta P}{\sqrt{e^{-t}(e^{2P} (\partial_t Q)^2 + (\partial_t P)^2 )}}\, d\theta.
	\end{align}	
From here on, the general expressions become extremely large and complicated, hence we will limit ourselves to the general behavior and in the appendix we provide the complete expressions. The connection one-form is
	\beq \label{conngowdy}
	\omega^1_{\ 2\, a} \Theta^a = \frac{1}{2}\frac{e^{t/2}}{\left( e^{2P} Q_t^2 + P_t^2\right)^{3/2} (P_\theta Q_t - P_t Q_\theta) } \bigg( U(t,\theta) \, \Theta^1 - V(t,\theta)\, \Theta^2 \bigg),
	\eeq
where we introduce the notation $P_t \equiv \partial_t P$ and $U$ and $V$ are expressions in terms of $P$, $Q$ and their derivatives with respect to $t$ and $\theta$. The curvature two-form has the general expression
	\beq  \label{curvgowdy}
	R^1_{\ 2ab}\, \Theta^a \wedge \Theta^b = - \frac{e^t}{(e^{2P} Q_t^2 + P_t^2)^3 (P_\theta Q_t - P_t Q_\theta)^3} W(t,\theta)\, \Theta^1 \wedge \Theta^2,
	\eeq
from where we obtain the Ricci scalar
	\beq  \label{riccigowdy}
	R =- \frac{e^{t}}{(P_t Q_\theta - Q_t P_\theta)^3} \tilde{W}(t,\theta),
	\eeq
where $W$ and $\tilde{W}$ are expressions in terms of $P$, $Q$ and their derivatives with respect to $t$ and $\theta$. From the expressions above we see that the connection one-form \eqref{conngowdy} is well defined and regular everywhere, except at the initial singularity $t \to \infty$, as we can observe form (\ref{curvgowdy}, \ref{riccigowdy}). Therefore, also in this case \textit{there are not quantization conditions to impose}. Using a particular solution for the case of asymptotically velocity term dominated cosmologies we can easily visualize the behavior of this solution. Consider the following choice of functions for (\ref{avtdP}, \ref{avtdQ}): $A(\theta)=1$, $B(\theta)=0$, $C(\theta) = 1$ and $D(\theta) = \theta$, then the induced metric is
	\beq   \label{indmetGowdyAVDT}
	h = \frac{1}{2}e^{-t}\, dt \otimes dt + \frac{1}{2}e^t\, d\theta \otimes d\theta,
	\eeq
while for the orthonormal basis and its dual we obtain 
	\beq   \label{framegowdyAVDT}
	e_1 = \sqrt{2} e^{t/2}\, \partial_t, \quad e_2 = \sqrt{2} e^{-t/2}\, \partial_\theta \quad \text{and} \quad \Theta^1 =\frac{1}{\sqrt{2}} e^{-t/2}\, dt, \quad \Theta^2 = \frac{1}{\sqrt{2}} e^{t/2}\, d\theta.
	\eeq
The connection one-form reduces to
	\beq  \label{conngowdyAVDT}
	\omega^1_{\ 2\,a}\, \Theta^a = -\frac{1}{\sqrt{2}} e^{t/2}\, \Theta^2,
	\eeq
and the curvature two-form and the Ricci scalar are
	\beq   \label{curvgowdyAVDT}
	R^1_{\ 2ab} \Theta^a \wedge \Theta^b = - e^t\, \Theta^1 \wedge \Theta^2,
	\eeq
	\beq   \label{riccigowdyAVDT}
	R = -2 e^t.
	\eeq
In this case we can observe in a simpler way that no discretization conditions arise. In fact, it is not difficult to show that the pfb is trivial calculating the Euler characteristic class $e(\mathcal{P})$ and computing the topological invariant from it. The topological invariant is obtained by the following expression \cite{nashsen}
	\beq  \label{eulernumber}
\frac{1}{2\pi} \left[ \sum_{i=1}^n (\pi - \alpha_i) + \int_{\partial \mathcal{M}} \kappa ds \right] + \int_{\mathcal{M}} e(\mathcal{P}) = \chi(\mathcal{M}),
	\eeq
where $\kappa$ is the geodesic curvature of the curves that conform the boundary of the manifold $\mathcal{M}$ and $\alpha_i$ the angles between any two pieces of the curves. The Euler form is given by $e(\mathcal{P}) = -1/(2\pi)\, R^1_{\ 2} = e^{t}/(4\pi)\, dt \wedge d\theta$ and we can choose curves at constant $t$ whose tangent vectors are $T = \sqrt{2} e^{-t/2} \partial_\theta$, then the geodesic curvature is $\kappa = 1/\sqrt{2} e^{t/2}$. These curves are closed and they do not have internal angles, hence integrating between $t_0$ and $t_1$ and for $\theta$ from $0$ to $\pi$ we obtain that $\chi(\mathcal{M}) = 0$. The Euler number does not depend on where these boundaries are set, then we can safely take the limits $t_0 \to -\infty$ and $t_1 \to \infty$. Therefore, we conclude \textit{no quantization conditions are imposed}, coinciding with the analysis previously done.

\subsubsection{The Einstein-Rosen gravitational waves}
\label{ssub:ERwaves}

Finally, we study the case of Einstein-Rosen gravitational waves. From \eqref{ERmetrics} and \eqref{induced} we obtain the induced metric for this case
	\begin{multline} \label{indER}
	h = \frac{\rho}{2} (e^{-4\psi} (\partial_t \Omega)^2 + 4 (\partial_t \psi)^2 )\, dt \otimes dt + \rho ( e^{-4\psi} \partial_t \Omega \partial_\rho \Omega + 4 \partial_t \psi \partial_\rho \psi)\, dt \otimes d\rho \\ 
	+ \frac{\rho}{2} ( e^{-4\psi} (\partial_\rho \Omega)^2 + 4 (\partial_\rho \psi)^2 )\, d\rho \otimes d\rho.
	\end{multline}
Again the metric is Lorentzian, then the reduced group of symmetries for the GHM is $SO(2)$. Therefore we introduce the orthonormal frame
	\begin{align}  \label{ERframe}
	e_1 &= \frac{\sqrt{2}}{\sqrt{\rho (e^{-4\psi} \Omega_t^2 + 4 \psi_t^2)}}\, \partial_t, \\
	e_2 &= \frac{ e^{2\psi} }{\sqrt{2} \sqrt{\rho (e^{-4\psi} \Omega_t^2 + 4 \psi_t^2)} (\psi_t \Omega_\rho - \Omega_t \psi_\rho)} \bigg((e^{-4\psi} \Omega_t \Omega_\rho + 4 \psi_t \psi_\rho)\, \partial_t + (e^{-4\psi} \Omega_t^2 + 4 \psi_t^2)\ \partial_\rho \bigg).
	\end{align}
and the dual frame
	\begin{align}  \label{ERdual}
	\Theta^1 &= \frac{\sqrt{\rho(e^{-4\psi} \Omega_t^2 + 4 \psi_t^2)}}{\sqrt{2}} \bigg(  dt + \frac{e^{-4\psi} \Omega_t \Omega_\rho + 4\psi_t \psi_\rho}{e^{-4\psi} \Omega_t^2 + 4 \psi_t^2}\, d\rho \bigg), \\
	\Theta^2 &= \frac{\sqrt{2\rho}\, e^{-2\psi} (\Omega_t \psi_\rho - \psi_t \Omega_\rho)}{\sqrt{e^{-4\psi} \Omega_t^2 + 4 \psi_t^2}}\, d\rho.
	\end{align}
The connection one-form is
	\beq \label{connER}
	\omega^1_{\ 2a} \Theta^a = \frac{1}{\sqrt{\rho} (e^{-4\psi} \Omega_t^2 + 4 \psi_t^2 )^{3/2}} \bigg( u(t,\rho) \Theta^1 + \frac{v(t,\rho)}{(\psi_t \Omega_\rho - \Omega_t \psi_\rho)}\, \Theta^2 \bigg), 
	\eeq
where $u$ and $v$ are expressions in terms of $\psi$, $\Omega$ and their derivatives with respect to $t$ and $\rho$. The curvature two-form is 
	\beq  \label{curvER}
	R^1_{\ 2 ab}\, \Theta^a \wedge \Theta^b = -\frac{2}{\rho}\, \Theta^1 \wedge \Theta^2,
	\eeq
and the Ricci scalar 
	\beq  \label{ricciER}
	R =  -\frac{4}{\rho}.
	\eeq
From the above expressions we notice that the connection one-form is regular except at those points where there is a curvature singularity $\rho = 0$, then it is found that also in this case \textit{there are not quantization conditions}, analogously to the previous cases.


\section{Final remarks}
\label{sec:remarks}

In this work we have presented the method of topological quantization applied to gravitational fields represented as generalized harmonic maps. 

The generalized harmonic maps introduced in \cite{Nettelghm} permit us to have a classical equivalent description of gravitational fields possessing two commuting Killing vector fields. In this way, the complete information about the system is included in the embeddeding of the base manifold into the target space and in the energy-momentum tensor of the generalized harmonic maps. The invariance under diffeomorphisms of the four-dimensional theory is identified as the invariance under reparametrizations of the embedded surface. 

The method of topological quantization can be applied once we have constructed the principal fiber bundle. As a result of the application of the method we obtain that for the three cases analyzed here, -i.e. Schwarzschild, unpolarized Gowdy $T^3$ and Einstein-Rosen gravitational waves-, we do not find any quantization condition to impose on the parameters that enter in the description of the gravitational configurations represented as generalized harmonic maps. This could be interpreted as a sign of the incompatibility of a classical description of the gravitational fields and the classical symmetries of spacetime with a quantum description of gravity. This is consistent with results obtained from other directions. In fact, as we mentioned in the introduction, different approaches point towards the deformed symmetries scenarios as a consequence of the quantum nature or discreteness of spacetime. Moreover, the results obtained here are in contrast with the previous results obtained the topological quantization of gauge fields, where the quantization conditions imply a discrete relation between the charges of the monopole and the charged particle. 

Our results on the one hand coincide with results obtained starting from the natural description in four dimensions of the gravitational fields from the Einstein-Hilbert action \cite{Quevedonew}, but on the other hand they do not coincide with previous findings in the case of Einstein-Rosen gravitational waves, where it was found in \cite{QuePat1} that the $C$-energy density per length unit defined along the symmetry axis at a fixed time possess a discrete relation, that is $E_c = - \ln n$ with $n$ an integer. In that case, the fact of $E_c$ being a negative quantity was interpreted as a sign of a non classical origin of this quantity. An alternative expression can be considered as $C$-energy density, giving as a result the discrete spectrum $E_c = 1 - n^2$ \cite{QuePat1}. Although it seems we have contradictory results, it must be considered that in \cite{QuePat1} it was assumed that the symmetry axis for the Einstein-Rose solution is free of singularities, whereas here we did not make such assumption. It will be interesting to understand what are the origin and implications of such differences between the two approaches. It will be also worth analyzing in future work the case of gravitational fields in the presence of matter fields in the GHM representation and compare with the results obtained in \cite{QuePat1}, where topological quantization with respect to the gauge symmetry is performed and leads to discrete relations between the charges of the gauge field and the parameters of the spacetime solution, e.g. in the case of the Reissner-Nordstr\"om solution where the charge of the $U(1)$ fields and the mass of the black hole are in a discrete relation.

\section*{Acknowledgements}
The author acknowledges support from CONACYT grant No. 207934. I am grateful to A. Bravetti and H. Quevedo for the useful comments on this work and on the last version of this paper. I would like to thank G. Amelino-Camelia and the Quantum Gravity group at Sapienza for the valuable discussions on deformed symmetries and quantum gravity and for the hospitality.

\appendix*
\section{}
\label{sec:appA}

In this appendix we present the complete expressions for the connection and the Ricci scalar curvature for the different cases. 

\subsection*{Unpolarized Gowdy $T^3$}

	\beq 
	\omega^1_{\ 2\, a} \Theta^a = \frac{1}{2}\frac{e^{t/2}}{\left( e^{2P} Q_t^2 + P_t^2\right)^{3/2} (P_\theta Q_t - P_t Q_\theta) } \bigg( U(t,\theta) \, \Theta^1 - V(t,\theta)\, \Theta^2 \bigg),
	\eeq
where
	\begin{multline}
	U(t, \theta) = \sqrt{2} \left( 2 Q_t^4 P_\theta e^{3P} - 2 Q_t^3 P_t Q_\theta e^{4P} + Q_t^3 Q_\theta e^{3P} + 4 Q_t^2 P_t^2 P_\theta e^P \right. \\ 
	- 4 Q_t P_t^3 Q_\theta e^P + Q_t^2 P_t P_{tt} e^P + Q_t P_t^2 Q_\theta e^P + 2 Q_t P_t P_\theta Q_{tt} e^P \\ 
	\left. + 2 Q_t P_t Q_\theta P_{tt} e^P - 2 P_t^2 Q_\theta Q_{tt} e^P + P_t^3 P_\theta \right)
	\end{multline}
	\begin{multline}
	V(t, \theta) = \sqrt{2} \left( 2 Q_t^3 P_{\theta t} e^{2P} - Q_t^3 P_\theta e^{2P} - 2 Q_t^2 P_t Q_{\theta t} e^{2P}  + Q_t^2 P_t Q_\theta e^{2P} - 2 Q_t^2 Q_\theta P_{tt} e^{2P} \right.  \\
	 + 2 Q_t P_t^3 P_\theta + 2 Q_t P_t Q_\theta Q_{tt} e^{2P} - 2 P_t^4 Q_\theta + 2 Q_t P_t^2 P_{\theta t} - Q_t P_t^2 P_\theta \\
	 \left. - 2 Q_t P_t P_\theta P_{tt} - 2 P_t^3 Q_{\theta t} + P_t^3 Q_\theta + 2 P_t^2 P_\theta Q_{tt} \right)
	\end{multline}
The Ricci scalar is
	\beq
	R =- \frac{2 e^{t}}{(P_t Q_\theta - Q_t P_\theta)^3} \tilde{W}(t,\theta),
	\eeq
where
	\begin{multline}
	\tilde{W}(t,\theta) = 2 Q_\theta^3 P_t^2 - P_t^2 Q_\theta^3 - 6 P_t^2 P_\theta Q_t Q_\theta^2 + P_t^2 P_{\theta \theta} Q_\theta e^{-2P} - P_t^2 P_\theta Q_{\theta \theta} e^{-2P} \\ 
	+ 2 P_t Q_\theta^2 Q_t P_\theta + 6 P_t Q_\theta P_\theta^2 Q_t^2 - 2 P_t Q_\theta P_\theta P_{\theta t} e^{-2P} + 2 P_t P_\theta^2 Q_{\theta t} e^{-2P} + P_{tt} Q_\theta^3 - 2 Q_\theta^2 Q_t P_{\theta t} \\
	 - Q_\theta^2 P_\theta Q_{tt} + 2 Q_\theta Q_t P_\theta Q_{\theta t} + Q_\theta P_\theta^2 P_{tt} e^{-2P} - Q_\theta P_{\theta}^2 Q_t^2 + Q_\theta Q_t^2 P_{\theta \theta} \\ 
	 - 2 P_\theta^3 Q_t^3 - Q_t^2 P_\theta Q_{\theta \theta} - P_\theta^3 Q_{tt} e^{-2P}.
	\end{multline}

\subsection*{Einstein-Rosen gravitational waves}

The connection one-form is 
	\beq  
	\omega^1_{\ 2a} \Theta^a = \frac{1}{\sqrt{\rho} (e^{-4\psi} \Omega_t^2 + 4 \psi_t^2 )^{3/2}} \bigg( u(t,\rho) \Theta^1 + \frac{v(t,\rho)}{(\psi_t \Omega_\rho - \Omega_t \psi_\rho)}\, \Theta^2 \bigg),
	\eeq
where 
	\beq
	u(t,\rho) = - \sqrt{2} e^{-2\psi} \left( e^{-4\psi} \Omega_t^3 + 8  \Omega_t \psi_t^2 - 2 \Omega_{tt} \psi_t + 2 \psi_{tt}  \Omega_t \right), 
	\eeq
and
	\begin{multline}
	v(t,\rho) = \sqrt{2} \left( \Omega_\rho \Omega_{tt} \Omega_t \psi_t e^{-4\psi} - \Omega_\rho \psi_{tt} \Omega_t^2 e^{-4\psi} + 8 \Omega_\rho \psi_t^4 - 8 \psi_\rho \Omega_t \psi_t^3 \right. \\
	\left.  + \Omega_t^3 \psi_{\rho t} - \Omega_t^2 \Omega_{\rho t} \psi_t e^{-4\psi} + 4 \psi_\rho \psi_t^2 \Omega_{tt} - 4 \psi_\rho \psi_t \psi_{tt} \Omega_t + 4 \psi_t^2 \psi_{\rho t} \Omega_t - 4 \psi_t^3 \Omega_{\rho t} \right).
	\end{multline}

\end{document}